\documentclass[sigconf]{acmart}
\AtBeginDocument{%
  }

\usepackage{color}
\usepackage{flushend}
\usepackage{graphicx}
\usepackage{booktabs}
\usepackage[htt]{hyphenat}
\usepackage[]{xcolor}
\usepackage{tabularx}
\usepackage{multirow}
\usepackage[inline]{enumitem}
\usepackage{xspace}
\usepackage{acronym}

\usepackage{adjustbox}
\usepackage{subcaption}
\usepackage{arydshln}
\usepackage{amsmath} 
\usepackage{anyfontsize}

\definecolor{midnightgreen}{rgb}{0.0, 0.29, 0.33}

\newcommand{\ourpipeline}{\texttt{CONE-RAG}\xspace}

\acrodef{CSA}{Conversational Search Agent}
\acrodef{CONE-RAG}{COnverstional Nugget-based Evaluation for RAG}
\acrodef{PTKB}{Personal Text Knowledge Base}
\acrodef{TREC}{TExt Retrieval Conference}
\acrodef{iKAT}{Interactive Knowledge Assistance Track}
\acrodef{CAsT}{Conversational Assistance Track}
\acrodef{NIST}{National Institute of Standards and Technology}
\acrodef{LLM}{Large Language Model}
\acrodef{NLI}{Natural Language Inference}
\acrodef{IR}{Information Retrieval}
\acrodef{RAG}{Retrieval Augmented Generation}

\newcommand{\header}[1]{\vspace{2mm}\noindent\textbf{#1}}

\setcopyright{acmlicensed}
\copyrightyear{2018}
\acmYear{2018}
\acmDOI{XXXXXXX.XXXXXXX}
\acmConference[Conference acronym 'XX]{Make sure to enter the correct
  conference title from your rights confirmation email}{June 03--05,
  2018}{Woodstock, NY}

\acmISBN{978-1-4503-XXXX-X/2018/06}

\begin{document}

\title{Conversational Gold: Evaluating Personalized Conversational Search System using Gold Nuggets}

\author{Zahra Abbasiantaeb}
\authornote{Equal contributions.}
\affiliation{%
  \institution{University of Amsterdam}
  \city{Amsterdam}
  \country{The Netherlands}
}
\email{z.abbasiantaeb@uva.nl}

\author{Simon Lupart}
\authornotemark[1] 
\affiliation{%
  \institution{University of Amsterdam}
  \city{Amsterdam}
  \country{The Netherlands}
}
\email{s.c.lupart@uva.nl}

\author{Leif Azzopardi}
\affiliation{%
  \institution{University of Strathclyde}
  \city{Glasgow}
  \country{Scotland, UK}
}
\email{leif.azzopardi@strath.ac.uk}

\author{Jeffery Dalton}
\affiliation{%
  \institution{University of Edinburgh}
  \city{Edinburgh}
  \country{Scotland, UK}
}
\email{jeff.dalton@ed.ac.uk}

    \author{Mohammad Aliannejadi}
\affiliation{%
  \institution{University of Amsterdam}
  \city{Amsterdam}
  \country{The Netherlands}
}
\email{m.aliannejadi@uva.nl}

\renewcommand{\shortauthors}{Abbasiantaeb et al}

\begin{abstract}

The rise of personalized conversational search systems has been driven by advancements in Large Language Models (LLMs), enabling these systems to retrieve and generate  answers for complex information needs. 
However, the automatic evaluation of responses generated by Retrieval Augmented Generation (RAG) systems remains an understudied challenge.  
In this paper, we introduce a new resource for assessing the retrieval effectiveness and relevance of response generated by RAG systems, using a nugget-based evaluation framework. 
Built upon the foundation of TREC iKAT 2023, our dataset extends to the TREC iKAT 2024 collection, which includes 17 conversations and 20,575 relevance passage assessments, together with 2,279 extracted gold nuggets, and 62 manually written gold answers from NIST assessors. 
While maintaining the core structure of its predecessor, this new collection enables a deeper exploration of generation tasks in conversational settings.  
Key improvements in iKAT 2024 include: 
(1) ``gold nuggets'' --- concise, essential pieces of information extracted from relevant passages of the collection --- which serve as a foundation for automatic response evaluation;
(2) manually written answers to provide a gold standard for response evaluation;
(3) unanswerable questions to evaluate model hallucination; 
(4) expanded user personas, providing richer contextual grounding; and 
(5) a transition from Personal Text Knowledge Base (PTKB) ranking to PTKB classification and selection. 
Built on this resource, we provide a framework for long-form answer generation evaluation, involving nuggets extraction and nuggets matching, linked to retrieval. 
This establishes a solid resource for advancing research in personalized conversational search and long-form answer generation. 
Our resources are publicly available at \url{https://github.com/irlabamsterdam/CONE-RAG}.
\end{abstract}



\maketitle

\section{Introduction}
Conversational information seeking provides a natural and intuitive way for users to interact and discover relevant information through dialogue with an agent~\cite{radlinski2017theoretical, azzopardi2018conceptualizing}. \acfp{LLM}~\cite{devlin2019bert} have taken us one step further to having access to \acfp{CSA} (e.g., ChatGPT, BingChat, Gemini, and BlenderBot). 
Moreover, wide-scale research and development in this area is more possible than ever, given the relative accessibility of the technology. 
However, despite advances and having numerous resources available for evaluating \acp{LLM} over a variety of tasks, \acp{CSA} present their own unique evaluation challenges~\cite{DBLP:conf/wsdm/Abbasiantaeb0KA24}. 
The complexity of \acp{CSA} stems from their interactive and nature~\cite{belkin2008iir}. Conversations can evolve in various ways with varied discourse~\cite{owoicho2023trec} and mixed initiatives~\cite{aliannejadi2019asking}. On top of that, conversational interactions are expected to be more personalized and tailored to the user than standard search systems~\cite{zamani2023conversational}. 

\acp{LLM} have greatly influenced how IR research is shaped~\cite{allan2024future}. 
Evaluation has been one of the main themes in the latest IR strategic meetings~\cite{allan2024future}, highlighting the impact of \acp{LLM} on the way we consume and evaluate information. This has led to increased community engagement in \acp{LLM}-based evaluation~\cite{faggioli2023perspectives,alaofi2024generative,thomas2024large,rau-etal-2024-bergen} leading to ongoing workshops like LLM4Eval~\cite{rahmani2024llm4eval,rahmani2024report} and Eval4RAG.\footnote{\url{https://eval4rag.github.io/}} Current research emphasizes on the importance of evaluating \ac{RAG} systems and its challenges~\cite{rau-etal-2024-bergen,samarinas2025icat}.

TREC Conversational Assistance Track (CAsT)~\cite{owoicho2023trec} ran for four years, aiming to provide a strong evaluation framework for conversational information-seeking tasks. 
TREC Interactive Knowledge Assistance Track (iKAT) 2023~\cite{aliannejadi2024trec} took a step further, providing complex decisional tasks that require multi-turn collaborative human-agent interactions. TREC iKAT 2023 also incorporated the personal knowledge graph in each dialogue, representing the noisy knowledge of the system about the user. 
Multiple TREC tacks in 2024 such as TREC RAG,\footnote{\url{https://trec-rag.github.io/}} iKAT,\footnote{\url{https://www.trecikat.com/}} and NeuCLIR~\footnote{\url{https://neuclir.github.io/}} focused on providing reusable \ac{RAG} collections, highlighting the importance and significance of this research direction.

\ac{RAG} evaluation faces several challenges as the nature of the task is complex and multi-step, which has resulted in a multitude of possible configurations~\cite{alaofi2024generative}. 
Given that the generated response of a \ac{RAG} system is a mix of the retrieved documents and the \ac{LLM}'s internal knowledge, existing surface-based QA metrics are not ideal for evaluation~\cite{rau-etal-2024-bergen}. 
Besides, existing \ac{RAG} collections offer more complex information needs, as opposed to ad-hoc retrieval and QA~\cite{samarinas2025icat}, calling for measuring the completeness of the generated responses. 
Existing research builds~\cite{samarinas2025icat,mayfield2024evaluation,takehi2023open,pradeep2024initial} on the ideas of nugget-based evaluation for summarization~\cite{giannakopoulos2013summary}, where the generated response is broken into information nuggets and compared against a set of gold nuggets for the information need. 
However, most prior work relies on LLM-generated nuggets, without having studied the effectiveness of LLM-based nugget extraction, potentially leading to unforeseen pitfalls and biases.

To address these limitations, in this work, we present experiment and extend the resources we created in TREC iKAT 2024. 
In particular, we present a set of complex multi-turn conversational topics, assessed for document relevance by the NIST assessors. 
We also collect human-extracted information nuggets from the relevant documents, along with a human crafted gold response (given the nuggets). 
We then conduct extensive experiments on the effectiveness of LLM-extracted nuggets in comparison with human nuggets. 
Furthermore, we conduct a crowd-sourced human nugget matching study to assess both different LLMs' nugget matching capabilities, as well as end-to-end RAG evaluation. 
Based on our experiments and collected data, we propose a novel extensible RAG evaluation framework, called \ourpipeline, which could effectively extract and match nuggets of a RAG response, and measure multiple nugget- and surface-based metrics. 
We publicly release conversational topics, document relevance assessments, human nuggets, gold responses, crowdsourced nugget matching labels, and \ourpipeline.
We believe that our experiments and provided resources will foster research in conversational RAG and nugget-based evaluation, as it provides useful insights on the problem.

\section{Related Work}

Development and evaluation of the interactive \ac{CSA} is an interesting perspective in \ac{IR}. 
The existing research~\cite{dalton2019cast,dalton2020cast,dalton2021cast,owoicho2023trec,over2001trec,DBLP:conf/trec/0001FS15,yang2016trec,DBLP:conf/trec/YangTS17} tried to facilitate the development of conversational search systems by proposing standard test collections.
The TREC Interactive Track (1998–2002)~\cite{over2001trec} and the TREC Dynamic Domain Track (2015–2017) \cite{DBLP:conf/trec/0001FS15,yang2016trec,DBLP:conf/trec/YangTS17} provided resources for passage retrieval across multiple rounds of feedback, focusing on iterative refinement rather than conversational interactions.
The TREC \ac{CAsT} \cite{dalton2019cast,dalton2020cast,dalton2021cast,owoicho2023trec,over2001trec,DBLP:conf/trec/0001FS15,yang2016trec,DBLP:conf/trec/YangTS17} was one of the first attempts to provide resource for conversational search task.
The track ran over four years resulting in TREC \ac{CAsT} 2019, 2020, 2021, and 2022 test collections.
The track evolved over four years by (1) making the conversations more complex, longer, and more dependent on the previous user--system interactions, (2) adding mixed-initiative interactions (clarification, feedback, elicitation, and etc) and (3) making the conversations multi-path based on different trajectories.  
These efforts resulted in more realistic and challenging conversational search scenarios.

The TREC \ac{iKAT}~\cite{aliannejadi2024trec} evolved the TREC \ac{CAsT} into a new track by making the conversations personalized. 
In personalized search given the same user query, the response of the system would be different to different users with different personas.
The TREC \ac{iKAT} 2023 enhanced the conversations with the persona of the users and added more complex information needs.
The persona of the user is provided as a set of natural language sentences and is static during the conversation.
The system needs to do reasoning over context, persona, and different sources of information to respond to the complex information needs.
The MTRAG collection focuses on the evaluation of multi-turn \ac{RAG} systems by providing manually collected and simulated topics. However, the dataset lacks a benchmark for passage retrieval.
In this work, we propose the TREC \ac{iKAT} 2024 collection which includes the resources for passage retrieval and response generation over personalized search.

Despite the advancements in \ac{RAG} systems and response generation task by the appearance of \acp{LLM}, the evaluation of the \ac{RAG} systems still remains a challenge.
The reference-based metrics such as Rouge~\cite{lin-2004-rouge} and BLEU~\cite{papineni-etal-2002-bleu} are commonly used to evaluate \ac{RAG} systems by measuring the overlap between generated responses and reference texts, assessing lexical similarity and relevance in terms of precision, recall, and n-gram matching.
One line of research focus on using \acp{LLM} for evaluation of \ac{RAG} systems~\cite{zhang-etal-2024-retrievalqa,es-etal-2024-ragas,kamalloo-etal-2023-evaluating,llmjudge2020,ares2024jon}. 
The BERGEN~\cite{rau-etal-2024-bergen} tool evaluates the RAG systems by measuring the reference-based and LLM-based metrics.
Another line of research~\cite{samarinas2025icat,pradeep2024initial} is attracted to the nugget-based evaluation of the RAG systems.
ICAT is an evaluation framework designed to assess the coverage of diverse factual information in long-form text generation. It decomposes the generated text into atomic claims, verifies each claim by retrieving information from a reliable knowledge source, and evaluates the alignment between these claims and the key aspects expected in the output.
Nugget-based evaluation has been first proposed in TREC Question Answering Track in 2003~\cite{DBLP:conf/trec/Voorhees03a}. The TREC 2024 Retrieval Augmented Generation (RAG) Track~\cite{pradeep2024initial} tries to automate nugget-based evaluation with the aid of \acp{LLM} to automatically extract and assign the nuggets.
In this work, we propose both manually collected and LLM-based nuggets of information for the TREC \ac{iKAT} 2024 collection. In addition, we propose an automatic pipeline for extracting the nuggets of information from the input text (can be either a response or a passage), matching them with the set of gold nuggets, and measuring the nugget recall and precision metrics for conversational search.

\section{Resources}
\label{sec:resources}
The goal of TREC \ac{iKAT} is to advance the evaluation of the personalized \acp{CSA}. To this aim, the track breaks the task of \acp{CSA} into three distinct components including (1) passage retrieval, (2) classification of the user persona, and (3) response generation.
In this work, we provide resources for different components of a \ac{CSA} by extending the existing resources from \cite{aliannejadi2024trec} and \cite{iKAT2024overview}, and by providing an automatic framework for evaluation, called \ourpipeline, of the quality of responses generated by RAG systems. In addition, we provide \ac{LLM}-based resources for different components of the personalized \ac{CSA}. 
We categorize and divide our resources into four groups, namely, (1) human-annotated resources collected by both NIST assessors and crowdsourcing, (2) resources collected by the aid of \acp{LLM}, (3) our new RAG evaluation pipeline (\ourpipeline), and (4) participants runs. 
We will further explain these resources in the following. \looseness=-1

\subsection{Human-Annotated Resources}
\header{Personalized conversational topics.} The TREC \ac{iKAT} 2024 collection includes 17 topics and each topic is associated with one distinct PTKB (persona). Different from \ac{iKAT} 2023 collection, one dialogue is developed for each topic to cover a larger array of topics and have a more diverse set of topics. In total, the dataset includes 218 user--system turns where each dialogue on average has 12.82 user--system turns. 
Each user--system turn of the conversation includes (1) user utterance, (2) resolved user utterance, (3) canonical response grounded on the passages from collection, (4) response provenance, and (5) PTKB provenance. 
The topics include longer and more complete responses compared to the TREC \ac{iKAT} 2023 collection.
The average length of canonical responses is 95.29 words while it is 77.26 for the \ac{iKAT} 2023 collection. 

For topic development, we mainly followed the same procedure and guidelines used for the development of the TREC \ac{iKAT} 2023 collection (see \cite{aliannejadi2024trec}) with some modifications.
During the topic development, we discarded the topics that were deemed too easy or too difficult, based on a preliminary GPT4 relevance assessment. 
We used the automatic relevance judgment model proposed by \cite{abbasiantaeb2024uselargelanguagemodels} for relevance assessment of personalized \acp{CSA}.

\header{Collection.} We use the existing collection from TREC \ac{iKAT} 2023 which is a subset of ClueWeb22-B~\cite{overwijk2022clueweb22} and includes 116,838,987 passages. 
We use the same segmentation code used in \ac{iKAT} 2023 for segmenting the documents. 
The \ac{iKAT} 2023 dataset included a Pyserini~\cite{PyseriniJimmyLin} index for the collection.
We extend the existing collection from \ac{iKAT} 2023 by publishing a newly learned sparse retrieval index. This index uses CoCondenser SelfDistil SPLADE++ checkpoint from HuggingFace\footnote{naver/splade-cocondenser-selfdistil}, using a numba index, as in the original SPLADE GitHub repository.

\header{Passage relevance assessment.} The relevance of the passages is assessed using the same scale of scores (0--4) used for TREC \ac{iKAT} 2023 and \ac{CAsT} collections. 
The NIST assessors judged the relevance of passages.
We selected a subset of 116 turns from 218 turns for relevance assessment by discarding the very general and clarification turns.
We tried to keep the turns with relevant PTKB statements for relevance assessment.

\header{Adaptive pooling.}
As the existing research~\cite{abbasiantaeb2024uselargelanguagemodels} has shown reusability issues for the TREC \ac{iKAT} 2023 dataset, we tried to mitigate this challenge by using an adaptive pooling approach rather than the existing static pooling approach. 
The adaptive pooling enabled us to assess up to the top 30 passages, leading to considerably more relevant passages.
To do so, we first leveraged automated GPT4o relevance assessment. The existing research~\cite{abbasiantaeb2024uselargelanguagemodels,faggioli2023perspectives} shows that \acp{LLM} tend to be more forgiving than human annotators (i.e., the average relevance scores are higher) while exhibiting a relatively low false negative rate. Based on these findings, we filtered out the passages in the top 30 pool and asked the NIST assessors to only assess the passages that were deemed relevant by GPT4o. 
To avoid reinforcing GPT4 biases and the LLM evaluation circularity problem, included all the top 5 passages in the final assessment pool.
But for the passages ranked between the top 5 and top 30, we applied the LLM filtering.
In doing so, the size of the final assessment pool is 20,575 passages and is judged by the NIST assessors.

\header{Gold response.} After assessment of the passage's relevance, We asked the NIST assessors to write a comprehensive gold response for the user utterances using the information from relevant passages. 
The NIST assessors provided a gold response for 62 turns, selected by the organizers based on their complexity.
Each gold response is written using a different number of relevant passages. An average number of 21.6 passages are used for writing the gold responses. 
The average length of the gold responses is approximately 104 words. \looseness=-1

\header{Nuggets of information.} The nuggets of information are extracted from relevant passages by NIST assessors. 
These nuggets serve as a resource for the evaluation of the response generated by RAG systems. 
Different from the existing approaches~\cite{pradeep2024initial,samarinas2025icat} that use a phrasal expression for the information nuggets, the nugget extracted by NIST assessors is a continuous span of the passages.
In total, 2,279 nuggets of information are extracted from 79 turns. 
More statistics about our collection of nuggets are provided in Table \ref{tab:stats-nuggets}.
As these nuggets are extracted from different relevant passages, there might be duplicates of information between them.
For example, however, the following two nuggets "Snake plants can survive with infrequent watering" and "Snake plants (Sansevieria) are highly drought-tolerant" are extracted from different passages, but they carry the same information and are considered duplicates.
We remove the duplicated nuggets and create a smaller set of nuggets.
After removing the duplicate nuggets, we have 1,201 nuggets.
We release both the original nuggets annotated by NIST assessors and the de-duplicated version of them.

\header{Nugget entailment labels.} 
Our goal is to benchmark the nugget entailment and matching. In particular, given a generated response, and an extracted information nugget, the task is to assess if the information nugget is entailed in the generated passage, i.e., is there a match? See Section~\ref{eval-our-rag} for more details. We release a total of 1,356 nugget entailment labels, collected via crowdsourcing on Amazing Mechanical Turk (MTurk). This resource will enable the researchers to develop and evaluate the entailment models for nugget-based RAG evaluation.

\header{PTKB statement relevance assessment.} The relevance of each PTKB statement to each conversational turn is assessed by the organizers during the topic development, as well as the NIST assessors. 
The PTKB statements are classified as relevant or irrelevant.
The organizers assessed the relevance of PTKB statements for all 218 turns in the dataset.
While the NIST assessors only judged the relevance of PTKB statements on the subset of turns that are selected for passage relevance assessment.
We release both sets of assessments.

\vspace{-8pt}

\subsection{Automatic Resources}

\header{\ac{LLM}-based relevance assessment pool.} We create a pool of 32,999 query--passage pairs by selecting the top 30 passages returned by participating runs. 
We use the GPT4o model and the code released by \cite{abbasiantaeb2024uselargelanguagemodels} to assess the relevance of passages in the pool.
The pool has 17,130, 8,771, 4,393, 1,993, and 712 passages with relevance scores of 0, 1, 2, 3, and 4, respectively.
We release this automatically judge pool as a resource.

\header{\ac{LLM}-extracted passage nuggets.} We use our nugget extraction model described in Section \ref{sec:nugg-extract} to extract the nuggets of information from relevant passages in the assessment pool, i.e., repeating what the NIST assessors did in extracting nuggets. 
Our model extracted in total of 6,680 nuggets for 79 turns while the human annotators extracted 2,279 nuggets from the same set of passages.
For more statistics on the nuggets by \ac{LLM} see Table \ref{tab:stats-nuggets}.

\header{\ac{LLM}-extracted response nuggets.} After extracting the passage nuggets in the previous step, we use the same model to extract nuggets from generated responses (submitted by iKAT participants) and release them as a resource.
It includes the nuggets by LLM for 26 different responses over 79 conversational turns.

\subsection{RAG Evaluation Pipeline.} We release the code for our RAG evaluation pipeline (\ourpipeline) as a resource. 
The researchers can use \ourpipeline to evaluate the quality of the generated response.
The released pipeline takes the generated responses for each user utterance as input in a JSON file and reports the following metrics:
\begin{enumerate}[nosep,leftmargin=*]
    \item Total nugget precision and recall based on four different sets of gold nuggets namely, human nuggets, de-duplicated human nuggets, LLM nuggets, and de-duplicated LLM nuggets.
    \item Nugget precision and recall over different turns from the dataset. 
    \item Rank of the submitted run compared to the baseline and participants' runs. 
    \item List of nuggets extracted by the input submission, with a label indicating whether it is matched with gold nuggets or not.
\end{enumerate}

\subsection{Participant Resources}

As an additional resource, we provide several submission runs, together with baselines for a total of 41 runs from 8 teams (including Organizer). 
The existing runs belong to three different categories including automatic (28 runs), manual (10 runs), and generation-only (3 runs).
In the manual runs, the retrieval and response generation models use both or one of the resolved\_utterance and relevant PTKB statements provided in the dataset.
The generation-only runs only provide the output of the response generation and use the provided ranking list of documents for response generation. 
25 of the run contain both retrieval and generation and 16 only retrieval.\footnote{A detailed description of the runs is provided in the overview of TREC iKAT 2024~\cite{iKAT2024overview}.}

\header{Retrieval.} For retrieval, most runs used a multi-step pipeline consisting of the following: 
(1) PTKB statement relevance prediction;
(2) conversational rewriting (most incorporating the previous canonical responses as well as predicted relevance PTKB statements) and conversational query expansion;
(3) retrieval using traditional lexical or neural IR models; and 
(4) multi-stage passage re-ranking with neural language models.

\header{Response generation.} For generation, the runs mostly relied on a RAG pipeline, using retrieved passages from the previous step with the conversation history or rewrite. Then, a diverse set of LLMs were used: Llama 8B and 70B, GPT4, GPT4o, and Gemini-1.5-flash.

\begin{table}[]
    \centering
    \caption{Statistics of Test Retrieval Data}
    \begin{tabular}{ll}
    \toprule
        Topics & \phantom{00,0}17 \\
        Turns & \phantom{00,}218 \\
        PTKB statements & \phantom{00,}288 \\
        Assessed topics & \phantom{00,0}14 \\
        Assessed turns & \phantom{00,}116 \\ 
        Avg.~dialogue length & \phantom{00,0}12.82 \\  
        Avg.~response length & \phantom{00,0} 95.29\\  
        Avg.~PTKB length & \phantom{00,0}16.94 \\        
        Passages assessed & 20,575 \\    
    \midrule
        Fails to meet (0) & 10,680\\
        Slightly meets (1) & \phantom{0}4,246\\
        Moderately meets (2) & \phantom{0}4,325\\
        Highly meets (3) & \phantom{0}1,199\\
        Fully meets (4) & \phantom{00,}125\\
    \midrule \midrule
        PTKB turns assessed by NIST & \phantom{00,}114 \\
        PTKB assessments by NIST & \phantom{0}1,917 \\
        Relevant (1) & \phantom{00,}201 \\
    \midrule
        PTKB turns assessed by the organizers & \phantom{00,}218 \\
        PTKB assessments by the organizers & \phantom{0}3,660 \\
        Relevant (1) & \phantom{00,}175 \\
    \bottomrule            
    \end{tabular}    
    \label{tab:stats-topics}
\end{table}

\begin{table}[]
    \centering
    \caption{Statistics of Answer Generation Data}
    \begin{tabular}{lll}
    \toprule
    & Assessed Turns &  79 \\
    \midrule
     \multirow{5}{*}{\rotatebox[origin=c]{90}{\textbf{Human}}}  & Number of Nuggets & 2,279 \\
     & Average number of nugget per turns & 28.84 \\
     & Average length of nuggets (word) & 32.36 \\
     & Number of Nuggets after removing duplicates & 1,201 \\
      \cmidrule{2-3}
     & Human-generated gold answer & 62 \\
     & Average length of gold answers & 21.6 \\
     \midrule
     \midrule
     \multirow{4}{*}{\rotatebox[origin=c]{90}{\textbf{LLM}}} & Number of Nuggets & 6,680 \\
     & Average number of nugget per turns & 84.55 \\
     & Average length of nuggets (word) & 13.66 \\
     & Number of Nuggets after removing duplicates & 3,760 \\
    \bottomrule            
    \end{tabular}    
    \label{tab:stats-nuggets}
\end{table}

\section{RAG Evaluation Pipeline Overview}
We propose \ourpipeline, a nugget-based evaluation pipeline for assessment of the generated responses.
The pipeline consists of two components namely nugget extraction and matching.
The nugget extraction component extracts the nuggets of the information from the input response. 
The matching component has two approaches to match either the response or the nuggets of the response to the gold nuggets to measure the nugget recall and precision.

\subsection{Nuggets Extraction} 
\label{sec:nugg-extract}
\begin{table}[!t]
\vspace{1em}
\caption{The prompt designed for nugget extraction model.}
\begin{tabularx}{\linewidth}{X}
\toprule
\# \textbf{Instruction:} \textit{I will give you a user query and a text to the user query. You should extract the nuggets of information related to the user query from the given text. The nuggets should be an exact copy of a span of text from the text. 
Please extract the nuggets and write each nugget in one line. If there is no nugget of information in the given text, please only say "No nugget". }\\

\# \textbf{User query:} \{$q_r$\} \\
\# \textbf{Text:} \{$t$\} \\
(Please copy exact spans from the text as nuggets)\\
\# \textbf{Nuggets:}  \\
\bottomrule
\end{tabularx}
\label{tbl:prompt-ref-extract}
\end{table}

We employ zero-shot prompting an LLM to generate the nuggets of the information from the input text given the user question.
The generated nuggets must be spans of the input text.
The prompt we use for nugget extraction is shown in Table \ref{tbl:prompt-ref-extract}.
The component extracts a set of nuggets called $\mathbb{N}_P$ from the input text called $t$. 
The $\mathbb{N}_P$ is a set of spans from the input text $t$, where each span is representative of one nugget and is shown as $n_p$.
The input text ($t$) can be a response ($R$) or a passage ($P$).
We use the resolved\_utterance ($q_r$) as the user query in the prompt as it is the self-contained query of the user containing both relevant statements from PTKB and the context of the conversation.
The nugget extraction function (called $Nuggetizer$) is shown in the following equation:
\begin{equation}
    \mathbb{N}_P = Nuggetizer(t, q_r)
\end{equation}

\subsection{Nugget Matching}
We employ two approaches for matching the input response with the gold nuggets: (1) by extracting nuggets from the input response $\mathbb{N}_P$ and matching individual nuggets with gold nuggets $\mathbb{N}_G$; and (2) by matching the input response $R$ directly to the gold nuggets $\mathbb{N}_G$, without extracting nuggets from the input response.

\header{Nugget to Nugget (NtN)}  
This approach assesses whether the extracted nugget entails a gold nugget or not. Given a set of extracted nuggets $\mathbb{N}_P$ and a set of gold nuggets $\mathbb{N}_G$, we compute entailment scores between all nugget pairs using the following function:
\begin{equation}
    s = \text{NtN}(n_p, n_g)
\end{equation}
where $n_p$ and $n_q$ are a single extracted nugget and a single gold nugget, respectively. 
$s=1$ if the extracted nugget $n_p$ entails the gold nugget $n_g$, and 0 otherwise. 
The matching process consists of iterating over all $(n_p, n_g)$ pairs of extracted nuggets and gold nuggets to determine coverage. We use a \ac{NLI} model to implement the NtN function.
This allows us to determine the subset of covered gold nuggets $\mathbb{N}_G^{\prime} \subseteq \mathbb{N}_G$, and the subset of input response nuggets that covers them $\mathbb{N}_P^{\prime} \subseteq \mathbb{N}_P$. As a result, we can compute both nugget recall and precision.

\begin{table}[!t]
\vspace{1em}
\caption{The prompt designed for NtR matching.}
\begin{tabularx}{\linewidth}{X}
\toprule
\# \textbf{Instruction:} \textit{ I will provide you with a response and a gold information piece. Your task is to determine whether the response captures this piece of information or not.}\\

\# \textbf{Gold Information:} \{$n_g$\} \\
\# \textbf{Response:} \{$R$\} \\
\# \textbf{Please answer the following}: \\
Does the Response capture the Gold Information? Only respond with ``yes'' or ``no'' without further explanation. \\
\# \textbf{Answer} (yes/no): \\

\bottomrule
\end{tabularx}
\label{tbl:prompt-nugget-matching}
\end{table}

\header{Nugget to Response (NtR)}  In this approach, we directly evaluate whether the generated response entails a gold nugget. Instead of comparing individual nuggets, the matching model is prompted with a response $R$ and a gold nugget $n_g$, and predicts a binary outcome. This method enables an assessment of whether a response sufficiently covers the gold nuggets without requiring extracting nuggets of the system input response.
We use the prompt shown in Table \ref{tbl:prompt-nugget-matching} for matching the input response $R$ with the gold nugget $n_g$. Note that we prompt the LLM with each gold nugget individually to break down the task into small units.
\begin{equation}
    s = \text{NtR}(R, n_g)
\end{equation}
We use zero-shot prompting on LLM to implement function NtR. 
This allows us to determine the subset of covered gold nuggets $\mathbb{N}_G^{\prime} \subseteq \mathbb{N}_G$ as the set of gold nuggets for which $s=1$.
Using this approach we can only compute the nugget recall.

\subsection{Nugget Duplicate Removal}
To ensure a more concise and non-redundant set of nuggets for evaluation, we perform \textit{de-duplication} by removing gold nuggets (or extracted nuggets) that can be entailed by another nugget within the same set. This process is applied separately to both human-annotated gold nuggets and LLM-generated gold nuggets.
For a given set of gold nuggets $\mathbb{N}_G$, we filter out any nugget $n_g \in \mathbb{N}_G$ if there exists another nugget $n_g' \in \mathbb{N}_G \setminus \{n_g\}$ such that $E(n_g', n_g) = 1$. $E(n_g', n_g)$ indicates that $n_g'$ entails $n_g$. Any nugget that is fully covered by another is removed, ensuring that only the most informative and distinct nuggets remain.
By eliminating redundant nuggets within each source, this deduplication step refines the evaluation set and prevents overestimation of recall.

\begin{table}[]
    \centering
        \caption{ Correlation between ranking of responses using gold nuggets ($\mathbb{N}_G$) extracted by LLM and human for NtN matching approach. The nugget set after removing duplicates are shown with [D]. The correlation is reported using Kendall's $\tau$ and Spearman's $\rho$ metrics. }
    \begin{tabular}{llccccc}
    \toprule
        \multirow{2}{*}{$\mathbb{N}_G$}     &  \multirow{2}{*}{$\mathbb{N}_G$}   & \multicolumn{2}{c}{Precision$_{\text{NtN}}$}&& \multicolumn{2}{c}{Recall$_{\text{NtN}}$} \\ \cmidrule{3-4} \cmidrule{6-7}
                  &           & $\tau$ & $\rho$       && $\tau$ & $\rho$  \\
                  \midrule
        Human     & Human [D] &  0.591 & 0.737 &  & 0.848 & 0.956   \\ 
        LLM       & LLM [D]   &  0.731 & 0.882 &  & 0.801 & 0.918   \\         
        Human     & LLM       &  0.649 & 0.814 &  & 0.731 & 0.889   \\ 
        Human [D] & LLM  [D]  &  0.649 & 0.853 &  & 0.661 & 0.832   \\ 
        Human     & LLM [D]   &  0.637 & 0.772 &  & 0.626 & 0.805   \\ 
        Human [D] & LLM       &  0.567 & 0.746 &  & 0.719 & 0.879   \\ 
        \bottomrule
    \end{tabular}

    \label{tab:nugget-corr}
\end{table}

\section{Evaluation}
In this section, we explain our evaluation method for three main tasks of TREC iKAT 2024 including response generation, passage retrieval, and PTKB statement classification. 

\subsection{Experimental Setup}

Our framework uses different models in the nugget extraction and nugget matching components. For the nuggets extraction, we rely on GPT4o~\cite{openai2024gpt4ocard} as the base LLM that we prompt in zero-shot. For the NtN matching part, we employ a \ac{NLI} DeBERTa~\cite{he2021deberta} model, \texttt{MoritzLaurer/DeBERTa-v3-base-mnli-fever-anli}. For the NtR, we compare both using LLMs and NLI models, comparing GPT4o and DeBERTa. Additionally, duplicate removal is also made with the same NLI DeBERTa model. We rely on the BERGEN GitHub for computing the surface base metrics.\footnote{https://github.com/naver/bergen}

\subsection{Proposed RAG Evaluation Pipeline}
\label{eval-our-rag}

\header{Nugget extraction.} We assess the performance of this component in two ways. 
\begin{enumerate}[nosep,leftmargin=*]
    \item First, we employ the nugget extraction model on the same set of relevant passages that are used for extracting nuggets by human. In this way, we create a new set of gold nuggets by LLM. We compare the extracted nuggets by LLM with the gold nuggets annotated by human.
    \item We apply the matching approaches using two different sources of gold nuggets including (1) gold nuggets from humans and (2) gold nuggets extracted by LLM (see Section \ref{sec:resources}). We rank the participating runs based on their performance using each set of gold nuggets. We compute the correlation between these two rankings.
\end{enumerate}

\header{Nuggets matching.} We run a human study to do the NtR matching task on Amazon Mechanical Turk (MTurk). 
We give a response with a gold nugget to the human annotators and ask them to answer the following question by selecting one of the ``Yes" or ``No" choices.
``Is the gold nugget of information covered in the response?''
We designed a comprehensive guideline for the task and to ensure the quality of annotations 1) we put a test question in a random location in the study and discarded the annotations of the users who gave a wrong answer to the test question, 2) limited the annotators to those who have more than 98\% approval rate, have successfully completed more than 5,000 Hits, and are from UK, US, or Australia. 
We randomly selected 25 turns from the 79 turns with manual nuggets and picked the best run for each participating team based on the passage retrieval performance.
In total, we had 1,356 pairs of response--nugget pairs and divided them into batches 136 batches where each batch includes 10 response--nugget pairs.
We included one additional test question in each batch of data and assigned each batch to 3 different annotators.
We select the choice with a majority vote between 3 annotators as the final answer.
We call the set of nuggets from gold nuggets matched by the input response by human $\mathbb{N}_P^{\prime}[H]$.
We compute the agreement between human assessors and our NtR model by comparing $\mathbb{N}_P^{\prime}$ and $\mathbb{N}_P^{\prime}[H]$.

\header{End-to-end matching.} We evaluate the end-to-end performance of our proposed model in the case of using the NtR matching model. 
We use the set of  $\mathbb{N}_P^{\prime}$ and $\mathbb{N}_P^{\prime}[H]$ to evaluate the participating runs~\footnote{Note that the gold nugget by human is used for the NtR}.
We rank the runs using each output and compute the rank correlation.

\begin{table}[]
    \centering
        \caption{Agreement between human and NtR matching.}
    \begin{tabular}{lcc}
    \toprule
        Model &  Accuracy & Cohen's $\kappa$ \\
        \midrule
        DeBERTa & 0.805 & 0.247 \\
         GPT4o& 0.90 & 0.610 \\
        \bottomrule
    \end{tabular}
    \label{tab:corr-ntr-human-llm}
\end{table}

\subsection{Response Generation}
To evaluate the quality of the generated responses by participating teams, we use two different evaluation paradigms including (1) surface-based metrics and (2) our proposed nugget-based RAG evaluation pipeline. 

\header{Reference-based evaluation.} We employ surface-based, semantic-based, and LLM-based metrics and compare our gold response as a reference with the input response. We report the metrics such as Rouge-1, Rouge-2, Rouge-L, BEM~\cite{bulian2022tomayto}, and LLMEval with both GPT-4o~\cite{hurst2024gpt} and SOLAR-10.7B-Instructv1.0~\cite{kim2023solar}. 
In addition, we report the groundedness metric~\cite{aliannejadi2024trec} which measures to what extent the input response is grounded to the top passages by the passage retrieval model.

\header{Nugget-based RAG evaluation pipeline.}
We employ our proposed RAG evaluation framework and get the set of gold nuggets covered by the generated response. We will explain in the following how can we calculate the recall and precision of two different approaches for matching.
\begin{itemize}[nosep,leftmargin=*]
\item{Nugget to Nugget.} We compute the nugget recall and precision in this approach using the following equations.

\begin{equation}
    \text{Recall}_{\text{NtN}} = \frac{|\mathbb{N}_G'|}{|\mathbb{N}_G|} = \frac{|\{ n_g \in \mathbb{N}_G \mid \exists n_p \in \mathbb{N}_P, \text{NtN}(n_p, n_g) = 1 \}|}{|\mathbb{N}_G|}
\end{equation}
\begin{equation}
    \text{Precision}_{\text{NtN}} = \frac{|\mathbb{N}_P'|}{|\mathbb{N}_P|} = \frac{|\{ n_p \in \mathbb{N}_P \mid \exists n_g \in \mathbb{N}_G, \text{NtN}(n_p, n_g) = 1 \}|}{|\mathbb{N}_P|}
\end{equation}
The former, Recall$_{\text{NtN}}$, measures the proportion of gold nuggets that are entailed by at least one extracted nugget. The latter, Precision$_{\text{NtN}}$, measures the proportion of extracted nuggets that entail at least one gold nugget.

\item{Nugget to response.} We define the nugget recall in this approach as follows.

\begin{equation}
    \text{Recall}_{\text{NtR}} = \frac{|\mathbb{N}_G''|}{|\mathbb{N}_G|}  = \frac{|\{ n_g \in \mathbb{N}_G \mid \text{NtR}(R, n_g) = 1 \}|}{|\mathbb{N}_G|}
\end{equation}
This metric evaluates recall by considering whether the full response directly supports the gold nuggets, independent of explicit nugget extraction. 

\end{itemize}

\subsection{Passage Ranking}
We assess the submitted runs using different cutt\_offs over metrics like precision, recall, and nDCG. The primary evaluation metric is mean nDCG@5. 
We compute the average over all conversational turns with the same weight.

\begin{table}[]
    \centering
        \caption{Rank correlation between systems using NtN and NtR matching models.}
    \begin{tabular}{llccccc}
\toprule
\multirow{2}{*}{Metric} & \multirow{2}{*}{$\mathbb{N}_G$} & \multicolumn{2}{c}{Precision$_{\text{NtN}}$}  && \multicolumn{2}{c}{Recall$_{\text{NtN}}$}  \\
\cmidrule{3-4} \cmidrule{6-7}
& &$\tau$ & $\rho$ && $\tau$ & $\rho$ \\
\toprule
\multirow{2}{*}{Recall$_{\text{NtR}}$} & Human & 0.614 & 0.786 &  & 0.778 & 0.923   \\ 
& LLM & 0.626 & 0.781 &  & 0.778 & 0.925   \\  

\bottomrule
    \end{tabular}
    \label{tab:rank-corr-diff-matching}
\end{table}

\begin{table*}[]
    \centering
    \caption{The value of rank correlation between using \ourpipeline and different reference-based evaluation metrics.}
    \resizebox{0.95\textwidth}{!}{
    \begin{tabular}{llcccccccccccccc}
    \toprule

\multirow{3}{*}{Matching} & \multirow{3}{*}{$\mathbb{N}_G$} & \multicolumn{2}{c}{Rouge-1} & \multicolumn{2}{c}{Rouge-2} & \multicolumn{2}{c}{Rouge-L} & \multicolumn{4}{c}{LLMeval} & \multicolumn{2}{c}{BEM} & \multicolumn{2}{c}{Groundedness} \\
\cmidrule{9-12}
 &  & &  &  & &&& \multicolumn{2}{c}{SOLAR} & \multicolumn{2}{c}{GPT4o} &  & \\

\cmidrule{3-16}
 &  & $\tau$ & $\rho$ & $\tau$ & $\rho$ & $\tau$ & $\rho$ & $\tau$ & $\rho$ & $\tau$ & $\rho$ & $\tau$ & $\rho$ & $\tau$ & $\rho$ \\
 \midrule
 
\multirow{2}{*}{Recall$_{\text{NtN}}$} & LLM & 0.147 & 0.184 & 0.074 & 0.074 & 0.088 & 0.113 & 0.574 & 0.735 & 0.735 & 0.9 & 0.574 & 0.75 & -0.397 & -0.525 \\
 & Human & -0.088 & -0.11 & -0.103 & -0.123 & -0.088 & -0.13 & 0.603 & 0.775 & 0.441 & 0.6 & 0.691 & 0.86 & -0.25 & -0.277 \\
\midrule
\multirow{2}{*}{Recall$_{\text{NtR}}$} & Human & -0.1 & -0.088 & -0.033 & -0.103 & -0.183 & -0.165 & 0.717 & 0.85 & 0.633 & 0.829 & 0.567 & 0.724 & -0.333 & -0.353 \\
  & LLM & -0.02 & -0.007 & -0.033 & -0.026 & -0.098 & -0.067 & 0.725 & 0.874 & 0.66 & 0.822 & 0.569 & 0.773 & -0.399 & -0.476 \\

\bottomrule
    \end{tabular}
    }
    \label{tab:corr-surface-vs-nugget}
\end{table*}

\section{Results \& Analysis}

\subsection{RAG evaluation: Comparison with human performance}

\header{LLM-based nugget extraction.} Human extracted 2,279 nuggets while our LLM-based model extracted 6,680 nuggets. 
The average length of the nuggets by humans is around 32 words while for nuggets by LLM is around 14 words. 
The LLM-based nuggetizer model tends to extract more fine-grained and specific nuggets. 
Among the 6,680 nuggets extracted by LLM, 144 of them exactly match and 4,189 of them partially match with human nuggets while 2491 of them are new nuggets. 
Among the nuggets that partially match 2,443 of them have overlap over at least two words, 1,746 of them overlap over one or two words, and 2,159 of them overlap over more than 4 words with the human nuggets.

\header{Rank correlation using LLM and human nuggets.} We evaluate the participating systems using \ourpipeline and rank them. 
We repeat this process using a different set of gold nuggets.
In Table \ref{tab:nugget-corr} we report the correlation between the ranking of participating systems using different sets of gold nuggets.
As can be seen, without removing the duplicate nuggets, the correlation between the ranking of systems based on nugget recall is 0.731 and 0.889 using Kendall's $\tau$ and Spearman's $\rho$ metrics, respectively.

\header{LLM vs.\ human for NtR matching.} We compute the agreement between humans and our NtR model in Table~\ref{tab:corr-ntr-human-llm}.
The GPT4o model performs better than DeBERTa on the NtR matching task by achieving Cohen's $\kappa$ agreement of 0.61 and an Accuracy of 0.9.

\header{Human vs.\ LLM for Recall$_{\text{NtR}}$}. We use our end-to-end framework (\ourpipeline) to calculate the Recall$_{\text{NtR}}$ for the runs and sort them. Also, we use the human for NtR matching and calculating the Recall$_{\text{NtR}}$.
The \ourpipeline (based on GPT4o) achieves a rank correlation of 0.867 and 0.943 compared to humans on Kendall's $\tau$ and Spearman's $\rho$ rank correlation metrics. Using DeBERTa for NtR matching, Kendall's $\tau$ and Spearman's $\rho$ rank correlation metrics are 0.6 and 0.829.

\header{NtN vs.\ NtR matching.}
The correlation between the ranking of systems using each method of matching in \ourpipeline is shown in Table \ref{tab:rank-corr-diff-matching}.
As can be seen, we achieve Kendall's $\tau$ and Spearman's $\rho$ rank correlation of 0.778 and 0.923 when we use these matching models and sort the systems.
Also, using the original set of gold nuggets by either LLM or humans, the correlation is approximately equal.
This observation shows the robustness of matching models against the input set of gold nuggets.

\header{Nugget-based evaluation vs reference-based evaluation.} We report the rank correlation between the ranking of systems using \ourpipeline  and reference-based metrics in Table~\ref{tab:corr-surface-vs-nugget}.
Surprisingly, we observe a very low correlation with the ROUGE, although this metric is also recall-oriented, using n-gram of predicted and human reference response.
This fact highlights the stark contrast in performance between these two evaluation paradigms where each method focuses on distinctly different aspects.
The reference-based metrics like Rouge only consider the semantic similarity, however, the nugget-based evaluation works by comparing the nuggets of the information in the content of the response.
Interestingly, the correlation between nugget recall and precision with groundedness is negative.  
A response can have a high nugget recall while the nuggets of information in the response are from the intrinsic knowledge of the response generation model and not from the input passages.
The \ourpipeline has a higher rank correlation with LLMEval and BEM metrics.

\begin{table*}[]
    \centering
        \caption{Response generation performance of the participants over different metrics.}
    \begin{tabular}{lccccccccccc}
    \toprule
\multirow{3}{*}{Run name} & \multicolumn{4}{c}{NtN} & \multicolumn{2}{c}{NtR} \\
\cmidrule{2-7} 
                          & \multicolumn{2}{c}{Human} & \multicolumn{2}{c}{LLM} & Human  & LLM  & Rouge-L & \multicolumn{2}{c}{LLMeval}  & BEM & Groundedness\\ 
  \cmidrule{2-7} \cmidrule{9-10}                       
                        & Precision & Recall & Precision & Recall & Recall & Recall &  &  SOLAR &  GPT-4o &  &  \\
\toprule
\multicolumn{12}{c}{Automatic} \\
\midrule
uva-3 & 0.454 & 0.138 & 0.504 & 0.120  & 0.368 & 0.379 & 0.191 & 0.952 & 0.629 & 0.252 & 0.387 \\
orga-3 & 0.470 & 0.128 & 0.532 & 0.121 & 0.368 & 0.379 & 0.200 & 0.984 & 0.629 & 0.283 & 0.468 \\
uva-2 & 0.458 & 0.125 & 0.565 & 0.141 & 0.339 & 0.390  & 0.199 & 0.935 & 0.79 & 0.265 & 0.435 \\
uva-1 & 0.461 & 0.111 & 0.568 & 0.146 & 0.367 & 0.364 & 0.199 & 0.935 & 0.71 & 0.269 & 0.355 \\
uva-4 & 0.449 & 0.110  & 0.543 & 0.130  & 0.395 & 0.425 & 0.197 & 0.984 & 0.758 & 0.272 & 0.339\\
nii-1 & 0.460 & 0.110 & 0.540 & 0.129 & 0.364 & 0.391 & 0.202 & 0.952 & 0.677 & 0.263 & 0.871\\
orga-5 & 0.424 & 0.108 & 0.504 & 0.115 & 0.342 & 0.354 & 0.197 &0.968 & 0.710 &  0.287 & 0.548  \\
orga-6 & 0.433 & 0.107 & 0.496 & 0.123 & 0.346 & 0.357 & 0.193 & 0.919 & 0.645 & 0.253 & 0.484\\
orga-1 & 0.457 & 0.104 & 0.465 & 0.098 & 0.358 & 0.325 & 0.184 & 0.823 & 0.548 & 0.267 & 0.710\\ 
rali-3 & 0.423 & 0.095 & 0.532 & 0.117 & 0.276 & 0.332 & 0.214 & 0.887 & 0.710 & 0.235 & 0.613\\ 
orga-4 & 0.436 & 0.092 & 0.507 & 0.107 & 0.301 & 0.333 & 0.195 & 0.935 & 0.597 & 0.267 & 0.532\\ 
rali-2 & 0.422 & 0.088 & 0.493 & 0.115 & 0.246 & 0.313 & 0.222 & 0.919 & 0.645 & 0.246 & 0.565 \\ 
orga-2 & 0.378 & 0.076 & 0.444 & 0.101 & 0.273 & 0.244 & 0.198 & 0.613 & 0.419 & 0.209 & 0.677\\
infos-2 & 0.443 & 0.067 & 0.540 & 0.107 & 0.149 & 0.180 & 0.237 & 0.803 & 0.645 & 0.253 & 0.097\\
infos-4 & 0.430 & 0.066 & 0.480 & 0.058 & 0.163 & 0.176 & 0.218 & 0.629 & 0.403 & 0.213 & 0.565\\ 
infos-1 & 0.378 & 0.061 & 0.450 & 0.068 & 0.155 & 0.169 & 0.228 & 0.787 & 0.581 & 0.225 & 0.290\\ 
infos-3 & 0.417 & 0.059 & 0.434 & 0.062 & 0.171 & 0.184 & 0.217 & 0.661 & 0.306 & 0.227 & 0.645\\
iires-1 & 0.378 & 0.022 & 0.325 & 0.020 & 0.048 & 0.048 & 0.091 & 0.016 & 0.032 & 0.122 & 0.855\\
ksu-1 & 0.293 & 0.018 & 0.178 & 0.007 & 0.043 & 0.041 & 0.143 & 0.065 & 0.065 & 0.148 & 0.750 \\
\midrule
\multicolumn{12}{c}{Manual} \\
\midrule
orga-8-m & 0.482 & 0.115 & 0.525 & 0.124 & 0.366 & 0.337 & 0.195  & 0.967 & 0.661 & 0.268 & 0.516 \\
uva-6-m & 0.512 & 0.149 & 0.556 & 0.144 & 0.393 & 0.413 & 0.198  & 0.983 & 0.790 & 0.283 & 0.419 \\
orga-7-m & 0.477 & 0.126 & 0.539 & 0.143 & 0.383 & 0.393 & 0.199  & 0.935 & 0.709 & 0.266 & 0.435\\
uva-5-m & 0.470 & 0.127 & 0.563 & 0.141 & 0.404 & 0.404 &  0.195  & 1.000 & 0.725 & 0.247 & 0.435 \\
\midrule
\multicolumn{12}{c}{Generation-only} \\
\midrule
nii-gen-only & 0.482 & 0.051& 0.468 & 0.062 & 0.237 & 0.258 &  0.174 & 0.580 & 0.435& 0.203& 0.919  \\
\bottomrule
    \end{tabular}
    \label{tab:result-pipeline}
\end{table*}

\begin{figure}
    \centering
    \includegraphics[width=0.9\linewidth]{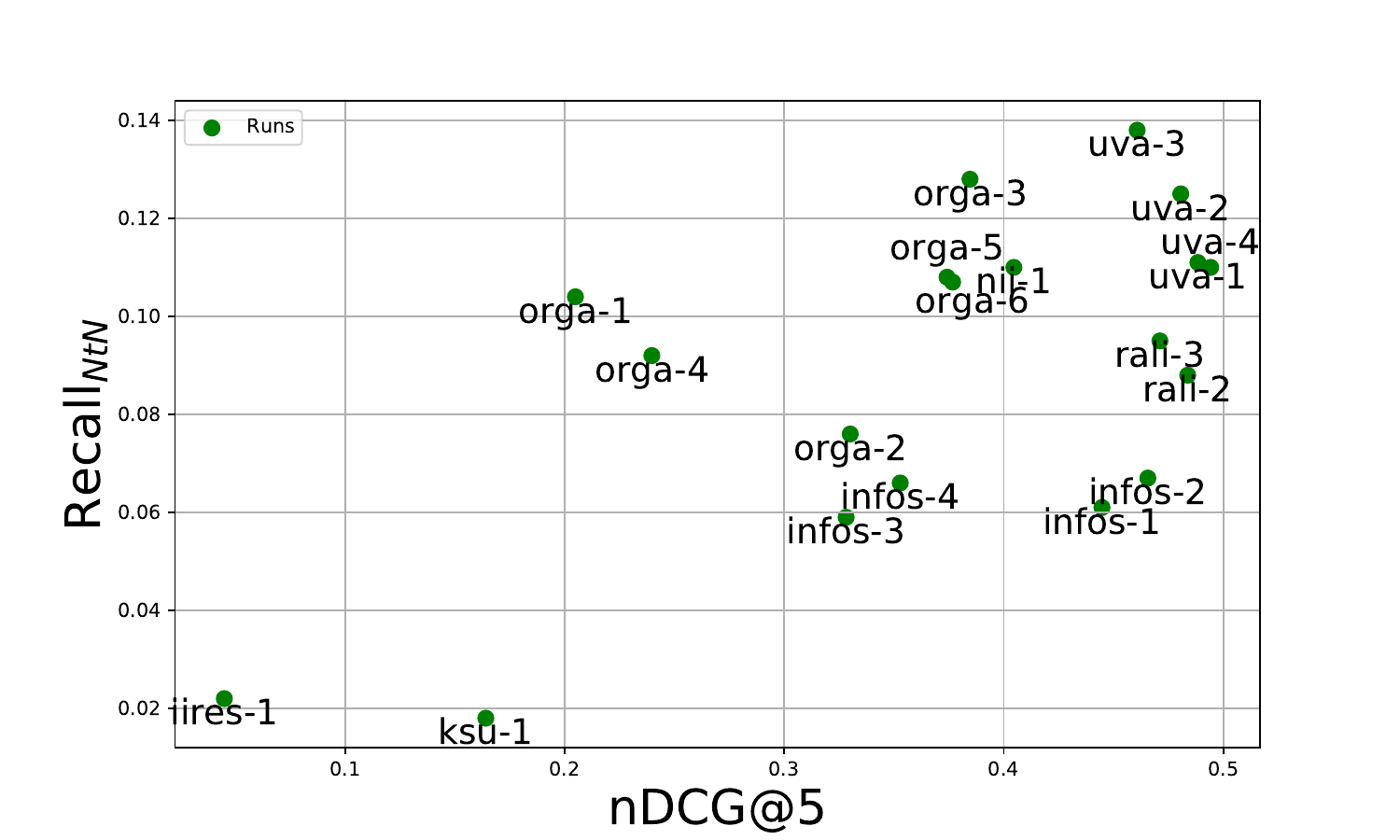}
    \caption{Nugget Recall vs.\ nDCG@5 for automatic runs.}
    \label{fig:nugget-recall-vs-ndcg}
\end{figure}

\header{Original nuggets vs.\ de-duplicated nuggets.} The correlation between the ranking of systems using original gold nuggets and after removing duplicate nuggets is shown in Table \ref{tab:nugget-corr}.
Removing the duplicate nuggets results in a higher decrease in correlation in over precision than recall. 
For example, after removing the duplicates from the original human gold nuggets, Kendall's Tau correlation is 0.591 and 0.848 over precision, and recall, respectively.

\header{Comparison between response generation performance of participating teams.} We report the performance of participating teams in Table~\ref{tab:result-pipeline}. 
Using the NtR matching model, the value of recall is much higher compared to the NtN matching model. 
This observation represents that the NtR matching model matches the response to a higher number of gold nuggets while the NtN model matches the lower number of extracted nuggets from the response to the gold nuggets.
Using the LLM or human nuggets as gold nuggets, we observe a lower difference in the value of recall over different teams.
However, the difference in the values of precision for each team is a bit higher compared to recall.
This is in line with the rank correlation in Table~\ref{tab:nugget-corr} where the rank correlation when using LLM or human gold nuggets is higher over the recall metric than the precision metric. 
Interestingly, the runs with higher nugget recall and precision have a lower groundedness value while teams with lower nugget recall and precision have a higher value of groundedness. 
For example, the ``iires-1'' team has a nugget recall value of 0.018 while it has a groundedness of 0.855. While the ``uva-3'' team has a nugget recall of 0.138 with a groundedness of 0.387.

\begin{table}[]
    \centering
        \caption{Average {Recall$_{\text{NtN}}$} and {Precision$_{\text{NtN}}$} of automatic runs over personal and non personal turns. }
    \begin{tabular}{llcc}
    \toprule
    Metric & $\mathbb{N}_G$ &  Personal  & Non Personal \\
    \toprule
\multirow{4}{*}{Recall$_{\text{NtN}}$} & Human & 0.064 & 0.115 \\
  & Human [D] & 0.021 & 0.049 \\
  & LLM & 0.066 & 0.135 \\
  & LLM [D] & 0.03 & 0.06 \\
  \midrule
 \multirow{4}{*}{Precision$_{\text{NtN}}$} & Human & 0.415 & 0.434 \\
  & Human [D] & 0.041 & 0.081 \\
  & LLM & 0.425 & 0.537 \\
  & LLM [D] & 0.152 & 0.201 \\
  \bottomrule
    \end{tabular}
    \label{tab:rag-perf-over-ptkb}
\end{table}

\begin{table}[t]
\centering
\caption{Automatic evaluation of passage retrieval results. Evaluation at retrieval cutoff of 1000.
}
 \label{tab:automatic-results}
  \resizebox{0.45\textwidth}{!}{%
    \begin{tabular}{lccccc}
\toprule
     Run &  nDCG@5 & P@20 & R@20 & R@1000 & mAP \\
\midrule
\multicolumn{5}{c}{Automatic}\\
\midrule
uva-4  & 0.494 & 0.596 & 0.176 & 0.783 & 0.355 \\
rali-2  & 0.484 & 0.478 & 0.148 & 0.580 & 0.198 \\
uva-2  & 0.481 & 0.549 & 0.166 & 0.644 & 0.297 \\
uva-1  & 0.488 & 0.583 & 0.172 & 0.774 & 0.334 \\
rali-3  & 0.471 & 0.471 & 0.140 & 0.564 & 0.189 \\
infos-2  & 0.466 & 0.516 & 0.148 & 0.682 & 0.256 \\
uva-3  & 0.461 & 0.506 & 0.153 & 0.486 & 0.219 \\
infos-1  & 0.445 & 0.472 & 0.141 & 0.639 & 0.235 \\
nii-1  & 0.405 & 0.437 & 0.138 & 0.598 & 0.218 \\
orga-3  & 0.385 & 0.381 & 0.131 & 0.427 & 0.176 \\
orga-6  & 0.377 & 0.416 & 0.126 & 0.420 & 0.171 \\
orga-5  & 0.374 & 0.397 & 0.134 & 0.603 & 0.213 \\
rali-1  & 0.367 & 0.387 & 0.121 & 0.580 & 0.170 \\
nii-2  & 0.366 & 0.394 & 0.119 & 0.519 & 0.197 \\
nii-3  & 0.365 & 0.392 & 0.119 & 0.519 & 0.197 \\
infos-4  & 0.353 & 0.401 & 0.114 & 0.570 & 0.190 \\
rali-4  & 0.347 & 0.375 & 0.112 & 0.564 & 0.161 \\
orga-2  & 0.33 & 0.404 & 0.119 & 0.563 & 0.202 \\
infos-3  & 0.328 & 0.353 & 0.099 & 0.529 & 0.153 \\
orga-4  & 0.240 & 0.253 & 0.081 & 0.304 & 0.106 \\
orga-1  & 0.205 & 0.234 & 0.065 & 0.263 & 0.089 \\
ksu-1  & 0.164 & 0.071 & 0.022 & 0.022 & 0.017 \\
dcu-4  & 0.150 & 0.154 & 0.042 & 0.185 & 0.055 \\
\midrule
\multicolumn{5}{c}{Manual}\\
\midrule
uva-6-m  & 0.529 & 0.594 & 0.197 & 0.706 & 0.349 \\
uva-5-m  & 0.473 & 0.547 & 0.183 & 0.706 & 0.305 \\
nii-5-m  & 0.455 & 0.535 & 0.166 & 0.596 & 0.255 \\
nii-4-m  & 0.454 & 0.535 & 0.165 & 0.596 & 0.256 \\
orga-7-m  & 0.414 & 0.46 & 0.149 & 0.706 & 0.245 \\
orga-8-m  & 0.413 & 0.435 & 0.143 & 0.460 & 0.19 \\
rali-6-m  & 0.403 & 0.365 & 0.113 & 0.460 & 0.129 \\
rali-5-m  & 0.396 & 0.365 & 0.113 & 0.460 & 0.132 \\
dcu-5-m  & 0.226 & 0.222 & 0.080 & 0.221 & 0.080 \\
dcu-6-m  & 0.207 & 0.211 & 0.073 & 0.217 & 0.072 \\
\bottomrule
\end{tabular}

}
\end{table}

\header{Performance per personal turns.} We classify the turns as personal and non-personal where the turns with at least one relevant PTKB statement are considered as personal turns. Looking at Table~\ref{tab:rag-perf-over-ptkb}, we observe that nugget precision and recall over personal turns is generally lower compared to non-personal turns.

\subsection{Retrieval performance evaluation}

\header{Overall results.} We show the retrieval performance of automatic and manual runs in Table \ref{tab:automatic-results}.
We do not observe a very large gap between the performance of manual and automatic runs which shows the advancement of LLMs for context modeling and understanding the information need of the user in the context of the conversation.
For example, the best manual and automatic teams (i.e., ``uva-4'' and ``uva-6-m'') have a 0.035 difference in terms of nDCG@5 metric.

\header{Correlation between passage retrieval and response generation performance.} As can be seen in Table \ref{tab:corr-retrieval-rag}, the correlation between response generation and passage retrieval is higher when using gold nuggets from LLM compared to using nuggets by humans. 
In addition, by removing duplicates from human gold nuggets, the correlation increases over both nugget precision and nugget recall metrics.
However, removing duplicates from LLM gold nugget, the correlation increases only in terms the nugget recall metric.
We do not observe a high correlation between nugget recall/precision and ndcg@5.
Based on this observation, we cannot say that better performance in passage retrieval guarantees a higher nugget recall/precision on response generation.
This gap between retrieval performance and response generation performance can be due to multiple reasons such as the focus of retrieval models on one aspect or the usage of intrinsic knowledge of response generation models.
However, further in-depth analysis is required to determine the main reason.
We show the nugget recall of the automatic runs (based on the NtN matching approach) vs the nDCG@5 in Figure~\ref{fig:nugget-recall-vs-ndcg}. 
The ``uva-X`` runs have the highest retrieval performance and nugget recall. These runs are based on the MQ4CS model~\cite{abbasiantaeb2024generate} which breaks the information need of the user into multi-aspect queries.
This observation indicates the effectiveness of leveraging multi-aspect queries for generating a more complete answer that covers a more diverse set of information. In addition, these runs used the GPT-4 model for response generation.

\begin{table}[]
    \centering
        \caption{Correlation between the ranking of systems based on passage retrieval (measured by nDCG@5) and response generation performance.}
    \begin{tabular}{lcccc}
    \toprule
    \multirow{2}{*}{Gold nuggets ($\mathbb{N}_G$)} & \multicolumn{2}{c}{Recall$_{\text{NtN}}$} &  \multicolumn{2}{c}{Precision$_{\text{NtN}}$}  \\
    \cmidrule{2-5}
            & $\tau$ & $\rho$ & $\tau$ & $\rho$  \\
    \midrule
 Human      & 0.404 & 0.563 & 0.310 & 0.46  \\ 
 Human [D]  & 0.415 & 0.602 & 0.392 & 0.574  \\ 
 LLM        & 0.556 & 0.744 & 0.614 & 0.763  \\
 LLM[D]     & 0.661 & 0.837 & 0.556 & 0.725  \\ 
 \bottomrule
    \end{tabular}
    \label{tab:corr-retrieval-rag}
\end{table}

\section{Conclusion}

We introduce the iKAT 2024 collection and resources, building on the foundations established by TREC iKAT 2024. These resources empower researchers to evaluate personalized conversational search agents (CSA) across both passage retrieval and response generation tasks. A key innovation of our work is \ourpipeline, a nugget-based evaluation pipeline for retrieval-augmented generation (RAG), along with an assessment pool for passage retrieval. The \ourpipeline enables the assessment of generated responses using nugget recall and precision metrics. Extensive experiments compare the framework's effectiveness against human performance, demonstrating its robustness and reliability.
We also provide the gold nuggets and matching of nuggets with responses which can be used for the development of nugget extraction and nugget matching models.
In addition, we propose the gold responses for each conversational turn as a resource which is essential for surface-based evaluation.

\bibliographystyle{ACM-Reference-Format}
\bibliography{references}

\end{document}